\documentclass[%
 aip,
 amsmath,amssymb,
 reprint,%
]{revtex4-1}

\usepackage{graphicx}
\usepackage{dcolumn}
\usepackage{bm}

\usepackage[utf8]{inputenc}
\usepackage[T1]{fontenc}
\usepackage{mathptmx}
\usepackage{etoolbox}

\usepackage{subcaption}
\usepackage{gensymb}
\usepackage[hidelinks]{hyperref}

\makeatletter
\def\@email#1#2{%
 \endgroup
 \patchcmd{\titleblock@produce}
  {\frontmatter@RRAPformat}
  {\frontmatter@RRAPformat{\produce@RRAP{*#1\href{mailto:#2}{#2}}}\frontmatter@RRAPformat}
  {}{}
}%
\makeatother
\begin{document}

\preprint{AIP/123-QED}

\title{Dynamic Phase Transitions in Mean-Field Ginzburg--Landau Models: Conjugate Fields and Fourier-Mode Scaling}

\author{Yelyzaveta Satynska}
\affiliation{$^1$School of Health, Science, and Sustainability, Roanoke College, Salem, VA, 24153}
\email{ysatynska@mail.roanoke.edu}
\author{Daniel T. Robb} 
\affiliation{$^1$School of Health, Science, and Sustainability, Roanoke College, Salem, VA, 24153} 
\date{\today} 

\begin{abstract}
Dynamic phase transitions of periodically forced mean-field ferromagnets are often described by a single order parameter and a scalar conjugate field. Building from previous work, we show that, at the critical period \(P_c\) of the Mean-Field Ginzburg--Landau (MFGL) dynamics with energy \(F(m)=a m^{2}+b m^{4}-h m\), the correct conjugate field is the entire even-Fourier component part of the applied field. The correct order parameter is \(z_k=\sqrt{\bigl|\lvert m_k\rvert^{2}-\lvert m_{k,c}\rvert^{2}\bigr|}\), where \(m_k\) is the \(k^{\text{th}}\) Fourier component of the magnetization \(m(t)\), and \(m_{k,c}\) is the \(k^{\text{th}}\) Fourier component at the critical period. Using high-accuracy limit-cycle integration and Fourier analysis, we first confirm that, for periodic fields that contain only odd components, the symmetry-broken branch below \(P_c\) exhibits \(z_k \sim \varepsilon^{1/2}\) (computationally tested for modes \(k \le 30\)), where \(\varepsilon=(P_c-P)/P_c\). This provides strong enough evidence that the 1/2 scaling holds for all Fourier modes. We then find three robust facts: (1) Exactly at \(P_c\), adding a small perturbation composed of even Fourier components with an overall field multiplier \(h_{\mathrm{mult}}\) yields \(z_k \sim h_{\mathrm{mult}}^{1/3}\) across many \(k\). (2) Mode-resolved deviations obey a parity rule: \(|\delta m_{2n}| \propto h_{\mathrm{mult}}^{1/3}\) and \(|\delta m_{2n+1}| \propto h_{\mathrm{mult}}^{2/3}\). (3) These scalings persist in two MFGL models with higher-order nonlinearities.
\end{abstract}

\maketitle
\section{Introduction}
Dynamic phase transitions (DPTs) are a special type of phase transitions which occur when external forces that act on a system change rapidly, rather than gradually. Unlike equilibrium phase transitions, during which the system has sufficient time to settle into equilibrium, DPTs involve time-dependent external forces, e.g. an oscillating magnetic field, which prevent the system from fully going through relaxation dynamics.

DPTs in magnetic systems are an active research topic, with continued theoretical, computational, and experimental developments.\cite{2014,Hohenberg,QuintanaBerger2023PRL,RiegoVavassoriBerger2017,QuintanaBerger2021,VatanseverEtAl2024,RiegoBerger2015,BuendiaRikvold2017,RiegoVavassoriBerger2018,MarinRamirez2020,QuintanaBerger2020,QuintanaValderramaBerger2023, QuintanaBerger2024, m_equil} DPTs have been observed in ultrathin ferromagnetic films driven by oscillating magnetic fields, where dynamic hysteresis loops reveal the bifurcation from a single symmetric loop into two asymmetric branches.\cite{RiegoVavassoriBerger2017,MarinRamirez2020,QuintanaBerger2020,QuintanaBerger2023PRL,QuintanaValderramaBerger2023} This allows for extraction of a dynamic order parameter $Q$ and its critical scaling.
In particular, Quintana and Berger recently demonstrated that ultrathin uniaxial Co films exhibit a magnetic DPT whose dynamic critical exponents agree with those of the two-dimensional Ising model.\cite{QuintanaBerger2023PRL}
Complementary experiments with asymmetric driving fields have shown that the dynamic order parameter is controlled by a generalized conjugate field that depends on the even components of the applied field.\cite{QuintanaBerger2021,QuintanaBerger2024}
These results show that dynamic hysteresis and the Fourier components of the conjugate field are experimentally adjustable in thin ferromagnetic films, motivating theoretical studies of these effects in simple model systems.

In parallel, Mean-Field Ginzburg--Landau (MFGL) and kinetic Ising models have shown that magnetic DPTs can share critical exponents with their equilibrium counterparts.\cite{2014,RiegoBerger2015,BuendiaRikvold2017,RiegoVavassoriBerger2018}
The goal of this article is to use the MFGL model to extend this picture: we identify the even-Fourier-component part of the applied field as the proper conjugate field for the DPT, show that the correct dynamic order parameter is built from Fourier components of the magnetization, and determine how these Fourier modes scale near the critical point.

\section{Background}
\begin{figure}[h]
    \centering
    \includegraphics[scale=.3]{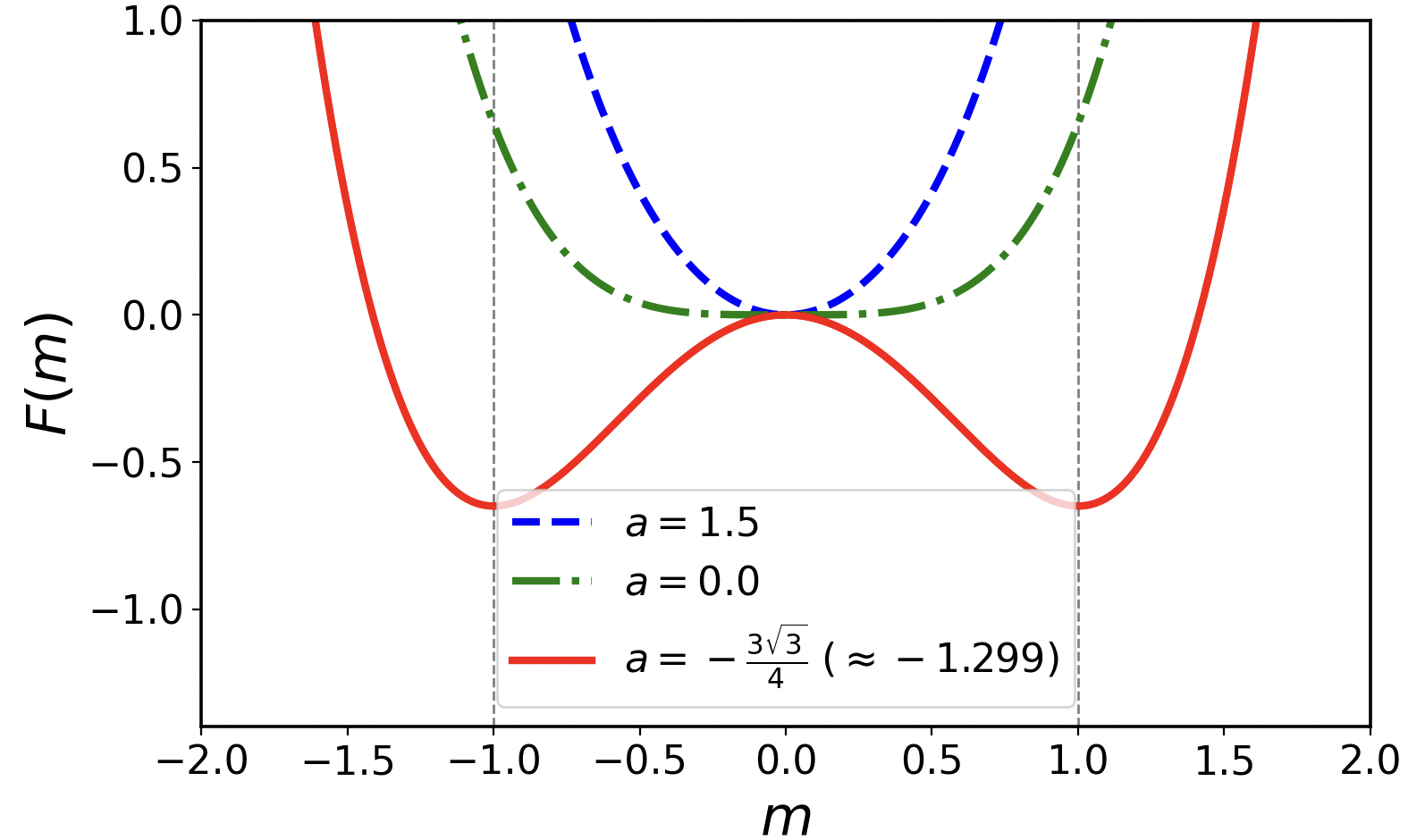}
    \caption{Ginzburg--Landau free energy $F(m)$ as a function of magnetization $m$ for $h=0$, $b=\tfrac{3\sqrt{3}}{8}$, and several values of $a$.}
    \label{fig1}
\end{figure}

Dynamic order parameter $Q$ is used to quantify magnetic DPTs. Prior to 2014, $Q$ was known as the time-averaged magnetization over one field period: $Q\approx0$ when the response follows the field symmetrically, and $Q\neq0$ when the symmetry is broken. Robb et al.\ (2014) showed that $Q$ is just one component of a more general dynamic order parameter.~\cite{2014} Their analysis was performed close to the \textit{critical period} $P_c$, defined such that $Q=0$ for  $P>P_c$, and $Q\neq0$ for $P<P_c$. They modeled the system with an MFGL free energy
$F(m)=am^2+bm^4-hm$, where $a$ and $b$ are material-specific constants, $m(t)$ is the magnetization, and $h(t)$ is the applied  field (see Figure~\ref{fig1}).

\vspace{-7px}
Robb et al.\ investigated the $a < 0$ double-well case when $F(m)$ has two local minima. We adapt their $a = -\frac{3\sqrt{3}}{4}$, $b = \frac{3\sqrt{3}}{8}$, which for $h = 0$ yield minima at $m = \pm 1$ (see Figure~\ref{fig1}).

\begin{figure}[htbp]
    \centering
    \includegraphics[scale=.3]{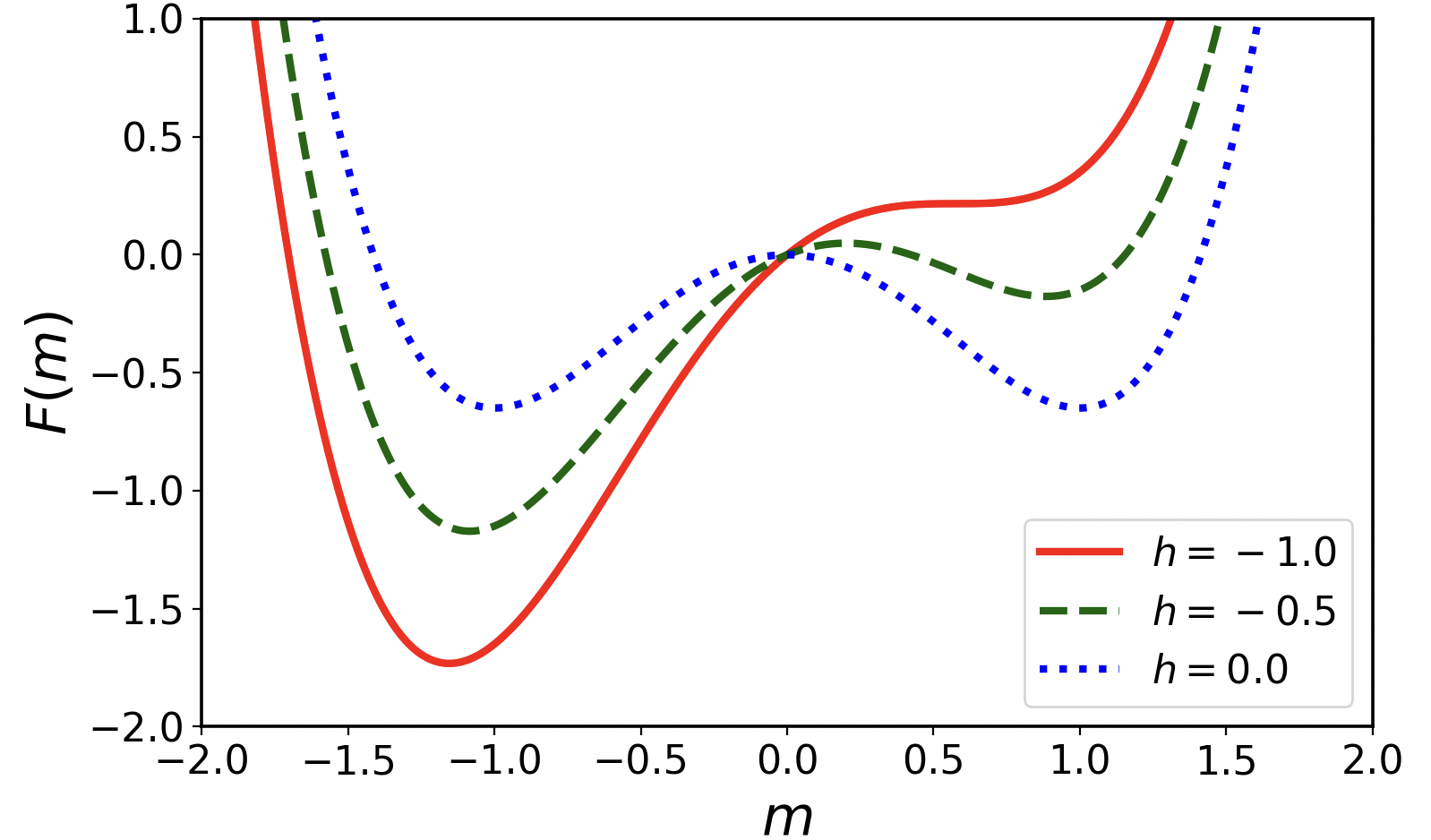}
    \caption{Ginzburg--Landau free energy $F(m)$ vs magnetization $m$ for different values of $h$ with $a=-\frac{3\sqrt{3}}{4}<0$ and $b = \frac{3\sqrt{3}}{8}$.\vspace{-8px}}
    \label{fig2}
\end{figure}

Figure~\ref{fig2} shows how a constant field $h$ tilts the double well. For $h=0$ the curve is symmetric, while negative $h$ deepens the minimum at negative $m$ and raises the one at positive $m$, 
favoring negative magnetization (case mirrored for positive $h$).
\vspace{-14px}

\begin{figure}[htbp]
  \centering
  \begin{subfigure}{0.468\textwidth}
    \centering
    \includegraphics[width=\linewidth]{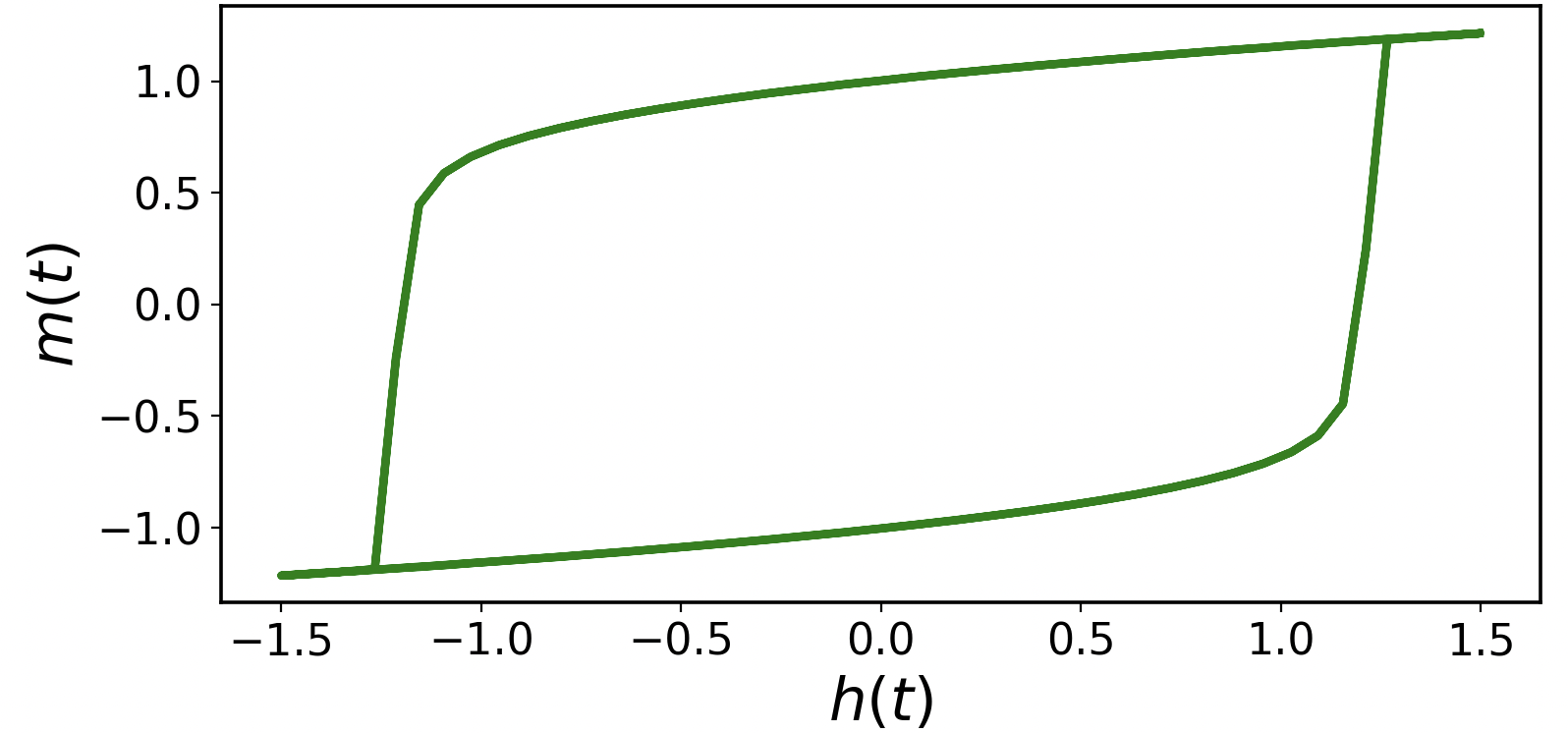}
    \caption{}
    \label{fig:left}
  \end{subfigure}
  \hfill 
  \begin{subfigure}{0.468\textwidth}
    \centering
    \includegraphics[width=\linewidth]{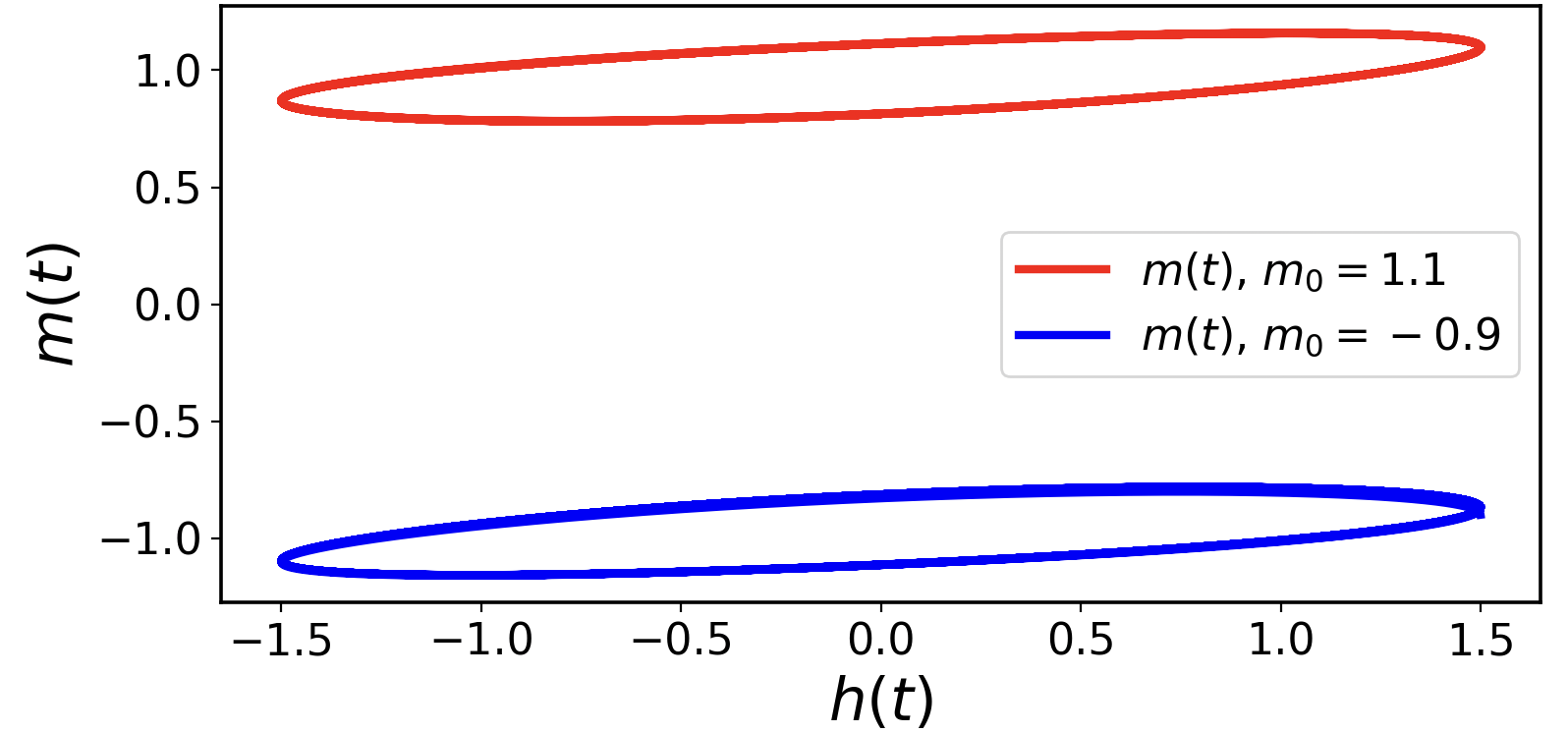}
    \caption{}
    \label{fig:right}
  \end{subfigure}
   \caption{Dynamic hysteresis in the MFGL model with critical period $P_c = 5.31935766199729$:
  (a) symmetric hysteresis loop for $P = 100 = 18.8 P_c \gg P_c$;
  (b) asymmetric hysteresis loops for $P = 1 = 0.188 P_c \ll P_c$.\vspace{-10px}}
  \label{fig:side-by-side}
\end{figure}

\vspace{-2px}
The magnetization relaxes to free-energy minima according to the Time-Dependent Ginzburg--Landau (TDGL) equation
\begin{equation}
  \frac{dm}{dt} = -\frac{dF}{dm} = -2am - 4bm^3 + h(t).
  \label{eq:tdgl}
\end{equation}
When $h(t)$ changes slowly (the oscillation period $P$ is large), the system will gradually fall from one well to the other exhibiting a symmetric hysteresis loop (Figure~\ref{fig:left}). For small $P$, the system does not have enough time to transition and gets stuck in one well exhibiting a nonzero period-averaged magnetization and two asymmetric hysteresis loops (Figure~\ref{fig:right}); the initial $m(0)$ determines which loop is followed and such separation is called a bifurcation of the DPT.

Because both $m(t)$ and $h(t)$ are periodic with period $P$, we can represent them as complex Fourier series with coefficients
\[
  m_k = \frac{1}{P} \int_0^P m(t)\,e^{-ikt}\,dt,
  \qquad
  h_k = \frac{1}{P} \int_0^P h(t)\,e^{-ikt}\,dt,
\]
so that $m_0$ and $h_0$ are the period-averaged magnetization and field. For equilibrium MFGL, the magnetization at the critical temperature $T_c$ scales as $m \propto h^{1/3}$.\cite{Hohenberg}
For the DPT in the same model, at $P=P_c$ the period-averaged magnetization obeys $m_0 \propto h_0^{1/3}$, so the exponent matches the equilibrium value.~\cite{2014,m_equil}

The critical period $P_c$ can be located via the loss of stability of the symmetric hysteresis loop. For a driving field with only odd Fourier components, the TDGL equation~\eqref{eq:tdgl} has a symmetric periodic solution $m(t)$ with $Q=m_0=0$. To test stability, we introduce a small perturbation $\delta m(t)$ via
$m(t)\to m(t)+\delta m(t)$ and keep only terms linear in $\delta m(t)$. Substituting into Eq.~\eqref{eq:tdgl} and subtracting the equation for $m(t)$,
\begin{equation}
  \frac{d}{dt}\,\delta m(t)
  = -\bigl[2a + 12b\,m^2(t)\bigr]\,\delta m(t)
  \label{eq:lin_delta_m}
\end{equation}
gives solution
\[
  \delta m(P)
  = \delta m(0)\,
    \exp\!\left(
      -\int_0^P [2a + 12b\,m^2(t')]\,dt'
    \right).
\]
The perturbation grows if the exponent is positive and decays if negative, so the stability boundary occurs when the exponent vanishes. Thus the critical period $P_c$ is determined by
\begin{equation}
  \int_0^{P_c} [2a + 12b\,m^2(t')]\,dt' = 0.
  \label{eq:Pc_integral}
\end{equation}

We located $P_c$ by numerically solving Eq.~\eqref{eq:Pc_integral}, which for our model gave $P_c = 5.31935766199729$, whereas Robb et al.\ located $P_c$ by the method below, finding $P_c = 5.319357661995$. At $P = P_c$ they expanded the symmetric solution as a series
$m(t) = \sum_k m_{k,c} e^{ik\omega_c t}$ with $\omega_c = 2\pi/P_c$. Using orthogonality,
\[
  \int_0^{P_c} m^2(t)\,dt
  = P_c \sum_k |m_{k,c}|^2.
\]
Substituting this into Eq.~\eqref{eq:Pc_integral} and dividing by $P_c$ they obtained the equivalent Fourier–space stability criterion
\begin{equation}
  \sum_k |m_{k,c}|^2 = -\,\frac{a}{6b},
\end{equation}
which they used to locate $P_c$.\cite{2014}

Robb et al.\ showed that the location of the critical period $P_c$ is determined by the odd Fourier components of $h(t)$. For $P \ge P_c$, the response to odd Fourier components of $h(t)$ consists solely of odd Fourier components of $m(t)$. For $P < P_c$, even components of $m(t)$ take on non-zero values and bifurcate from zero. At $P=P_c$, introducing even Fourier components of $h(t)$ couples the even and odd components of $m(t)$, and simulations for $j,k \le 30$ revealed the scaling relations
\begin{align}
\delta m_{k,e} &\propto (\delta h_{j,e})^{1/3}, &
\delta m_{k,e} &\propto (\delta h_{j,o})^{1/3}, \\
\delta m_{k,o} &\propto (\delta h_{j,e})^{1/3}, &
\delta m_{k,o} &\propto (\delta h_{j,o})^{1/3}.
\end{align}
where $e$ and $o$ denote even and odd Fourier components. This is a valuable discovery since the same scaling exponent $1/3$ is found for $m$ and $h$ in the equilibrium MFGL transition at  $T_c$.~\cite{Hohenberg}
\newpage 
\section{Reproduced Results}
\subsection{Order parameter definition and its scaling with $\varepsilon=\frac{P_c-P}{P_c}$}
\label{epsilon-scaling}
\begin{figure}[h]
    \centering
\includegraphics[scale=.3]{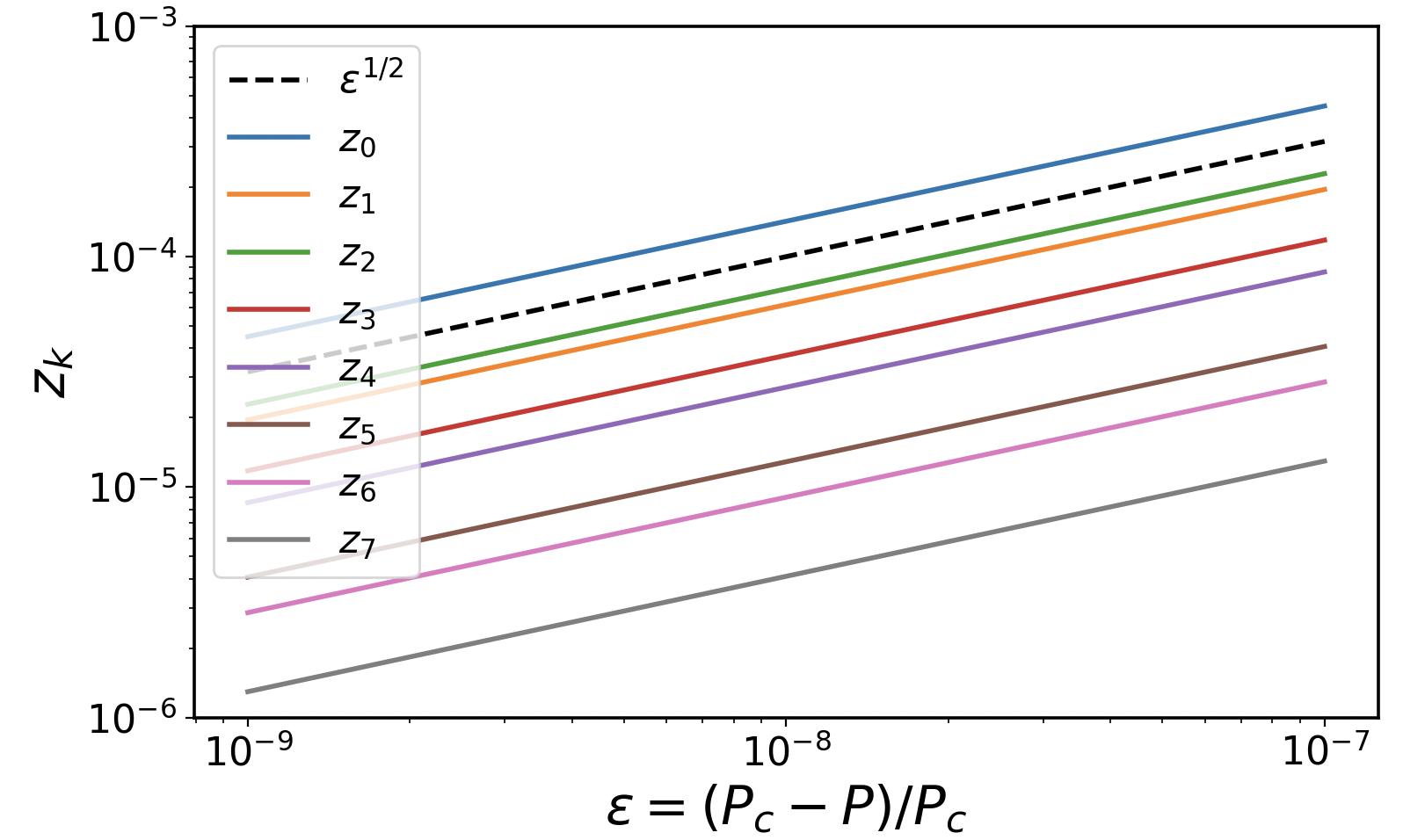}
    \caption{Scaling of the order parameter $z_k$ for $k=0,\dots,7$ vs.\ the
      scaled period $\varepsilon=(P_c-P)/P_c$ for $P<P_c$. The dotted line
      indicates the reference $\varepsilon^{1/2}$ scaling that the numerical
      data closely follow (re-created from Fig.~2 of Robb et al.\ 2014~\cite{2014}).}
    \label{fig5}
\end{figure}

\vspace{-5px}
Adapted from Robb et al., our
mode-resolved dynamic order parameter is
$
  z_k = \sqrt{\bigl||m_k|^2 - |m_{k,c}|^2\bigr|},
$
which measures the magnitude of the deviation of the $k$-th Fourier
component from the symmetric reference solution at $P_c$ (obtained for a
purely odd applied field).~\cite{2014} For a purely odd field, Robb et al.\ found that the symmetry-broken branch below $P_c$ obeys
\[
  z_k \propto \varepsilon^{1/2},
  \qquad
  \varepsilon = \frac{P_c-P}{P_c},
\]
matching the exponent in mean-field equilibrium.~\cite{Hohenberg}
Figure~\ref{fig5} shows our reproduction of this behavior for
$k \le 7$ at five subcritical periods $P<P_c$ (see Sec.~\ref{subsec:numerics} for precise values).
\vspace{-1px}
\vspace{-20px}
\subsection{Crossover from linear to cube-root scaling at $P=P_c$}
\begin{figure}[h]
    \centering
    \includegraphics[scale=.3]{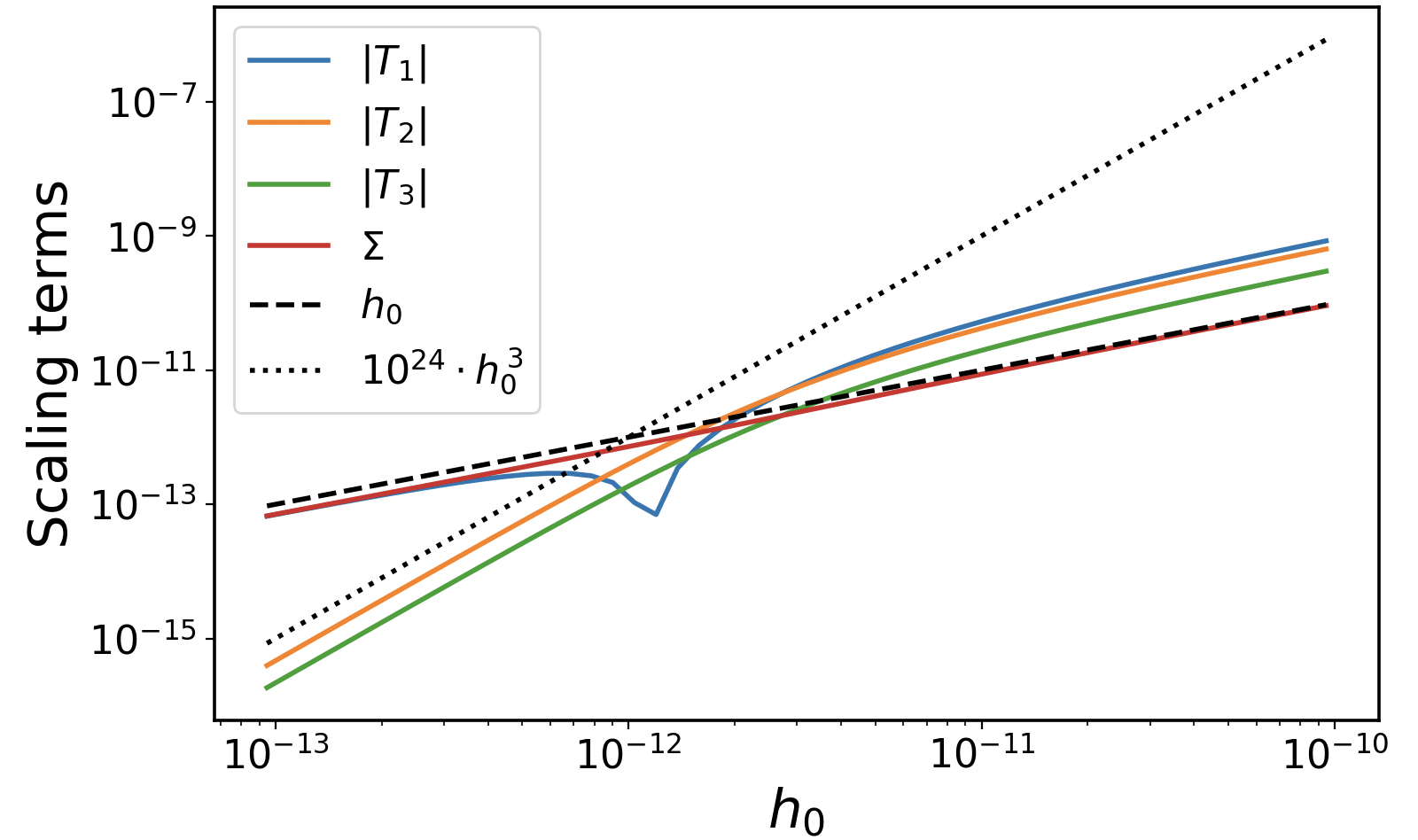}
    \caption{Scaling of $|T_1|$, $|T_2|$, $|T_3|$, and their sum $\Sigma$ vs.\ $h_0$, with a crossover near $h_0 \approx 10^{-12}$. Dashed and dotted lines show $\propto h_0$ and $\propto h_0^{3}$ scaling (re-created from Fig.~6 of Robb et al.~\cite{2014}).}
    \label{fig4}
\end{figure}

Reproducing Robb et al., we worked with small deviations
$m_k = m_{k,c} + \delta m_k$ and $h_k = h_{k,c} + \delta h_k$, inserted these
into the Fourier–mode TDGL equation (see Sec.~\ref{subsec:numerics} for details), subtracted the
critical solution, and grouped terms that are linear, quadratic, and cubic in
$\delta m_k$. This gave (cf.\ Eq.~(8) of Ref.~\onlinecite{2014})
\begin{equation}
  0
  = T_1(\delta m) + T_2(\delta m,\delta m) + T_3(\delta m,\delta m,\delta m)
    + \delta h_k,
  \label{eq:T123}
\end{equation}
with
\begin{align}
  &T_1(\delta m)
  = -2a\,\delta m_k
     - 12b \sum_{n_1,n_2}
       m_{n_1,c}\,m_{n_2,c}\,\delta m_{k-n_1-n_2},
  \label{eq:T1_def}\\
  &T_2(\delta m,\delta m)
  = -12b \sum_{n_1,n_2}
       m_{k-n_1-n_2,c}\,\delta m_{n_1}\,\delta m_{n_2},
  \label{eq:T2_def}\\
  &T_3(\delta m,\delta m,\delta m)
  = -4b \sum_{n_1,n_2}
       \delta m_{n_1}\,\delta m_{n_2}\,\delta m_{k-n_1-n_2}.
  \label{eq:T3_def}
\end{align}
Here $T_1$ is linear in $\delta m_k$, $T_2$ is quadratic, and $T_3$ is cubic. Figure~\ref{fig4} shows $|T_1|$, $|T_2|$, $|T_3|$, and their sum as functions of the $k=0$ Fourier component $h_0$ (a small constant offset added to the oscillating applied field). For the smallest $h_0$ the sum and $|T_1|$
follow the $\propto h_0$ line, while $|T_2|$ and $|T_3|$ are much smaller
and follow the $\propto h_0^3$ line. Around $h_0 \approx 10^{-12}$ the nonlinear terms grow comparable to $T_1$, showing how cubic terms balance the field near $P_c$, consistent with the cube--root scaling $|\delta m_0|\sim h_0^{1/3}$ reported in Figure~4 on page~4 of Reference~\onlinecite{2014}.

\vspace{-18px}
\section{New results} 
\subsection{Even--Fourier conjugate field and $z_k$ scaling}
\label{h_mult-scaling}
\begin{figure}[h]
    \centering
    \includegraphics[scale=.3]{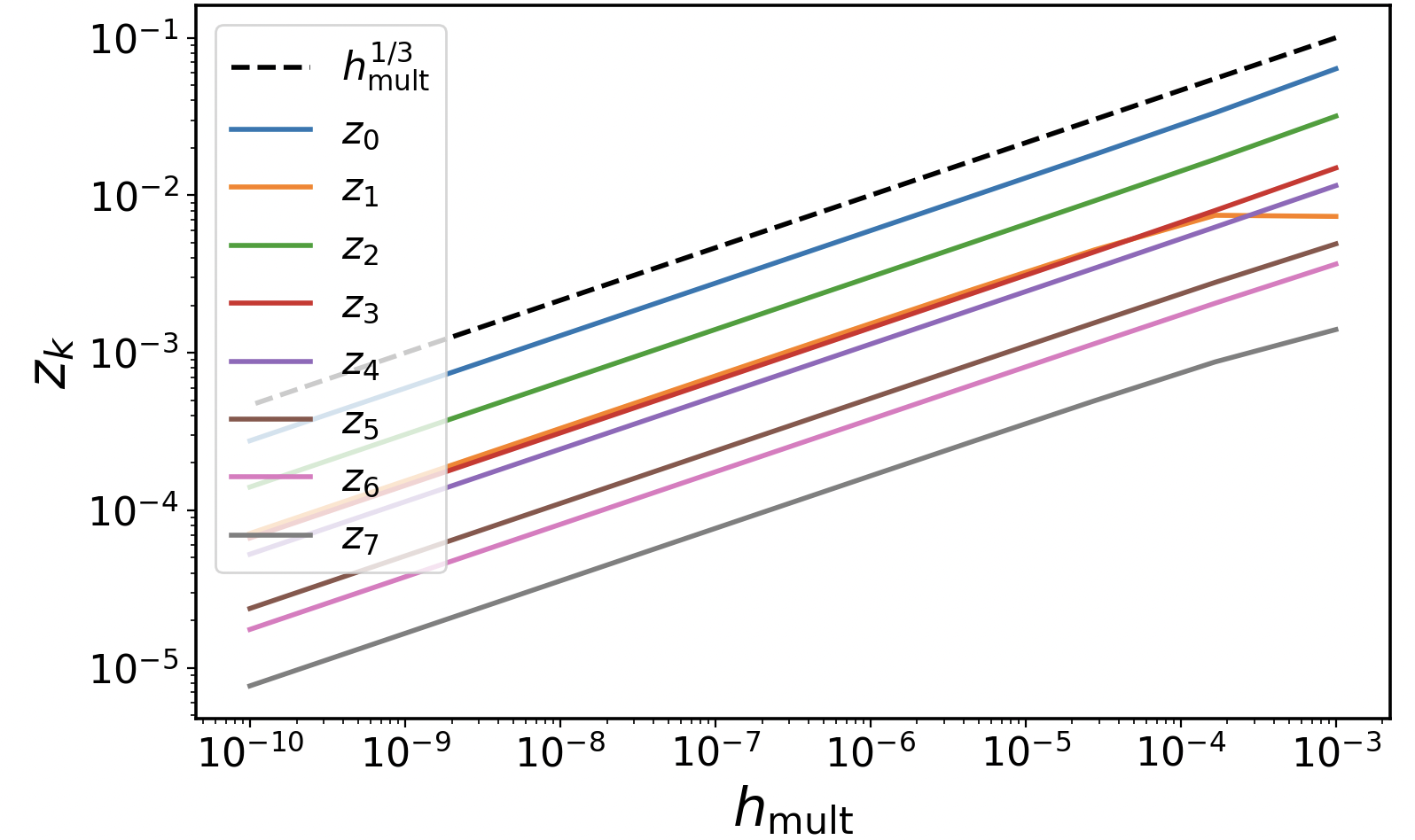}
    \caption{Scaling of the order parameter $z_k$ vs.\ conjugate field multiplier
$h_{\mathrm{mult}}$ at $P=P_c$. The dotted line shows the reference
$h_{\mathrm{mult}}^{1/3}$ scaling that the numerical data closely follow.}
\label{fig6}
\end{figure}

At $P = P_c$ we fixed the odd field part to $h_1\cos(2\pi t/P_c)$ with $h_1 = 1.5$ and added an even ``conjugate'' field part
\[
  h_{\text{even}}(t)
  = h_{\mathrm{mult}}\bigl[h_0
    + h_2\cos(2\cdot 2\pi t/P_c)
    + h_4\cos(4\cdot 2\pi t/P_c)\bigr],
\]
so that $h_{\mathrm{mult}}$ is a single scalar that scales the entire
even–Fourier field (see Sec.~\ref{subsec:numerics} for the coefficients $h_{\{0,2,4\}}$).  Varying $h_{\mathrm{mult}}$ as shown in Figure~\ref{fig6}, we found
\vspace{-10px}\[
  z_k \propto h_{\mathrm{mult}}^{1/3},
\]\vspace{-5px}
for $k\le 7$, with and without a constant offset $h_0$ in
$h_{\mathrm{even}}(t)$.
\vspace{-17.6px}
\subsection{Parity rule for mode deviations}
\begin{figure}[htbp]
  \centering
  \begin{subfigure}[b]{0.24\textwidth}
    \centering
    \includegraphics[width=\linewidth]{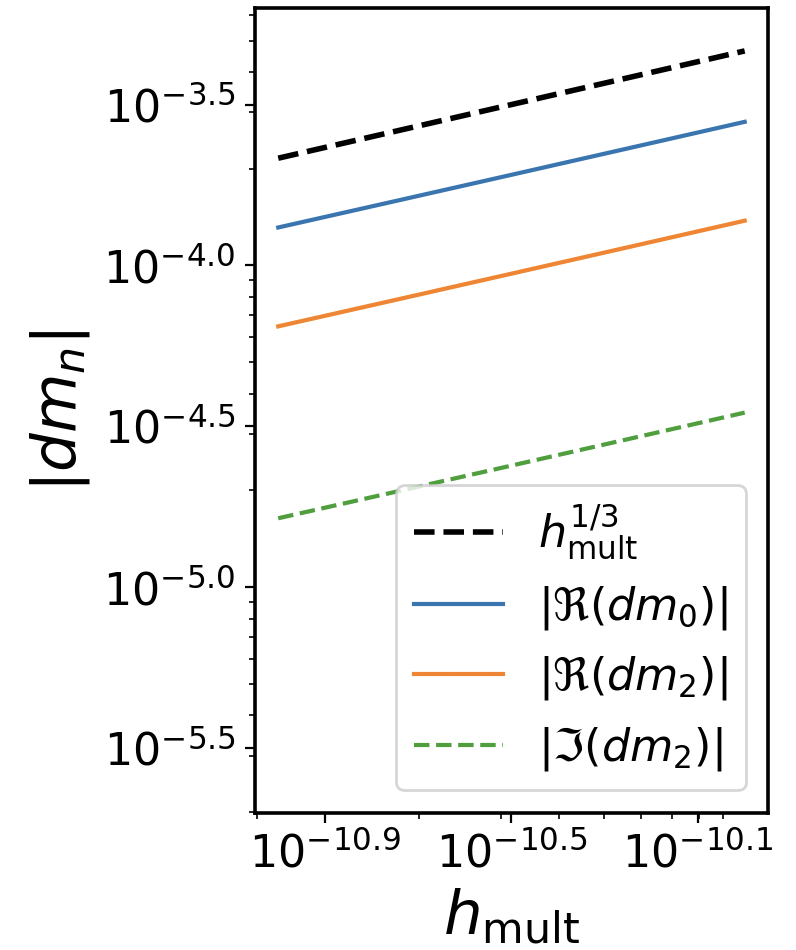}
    \caption{} 
    \label{fig:7a}
  \end{subfigure}%
  \hfill
  \begin{subfigure}[b]{0.24\textwidth}
    \centering
    \includegraphics[width=\linewidth]{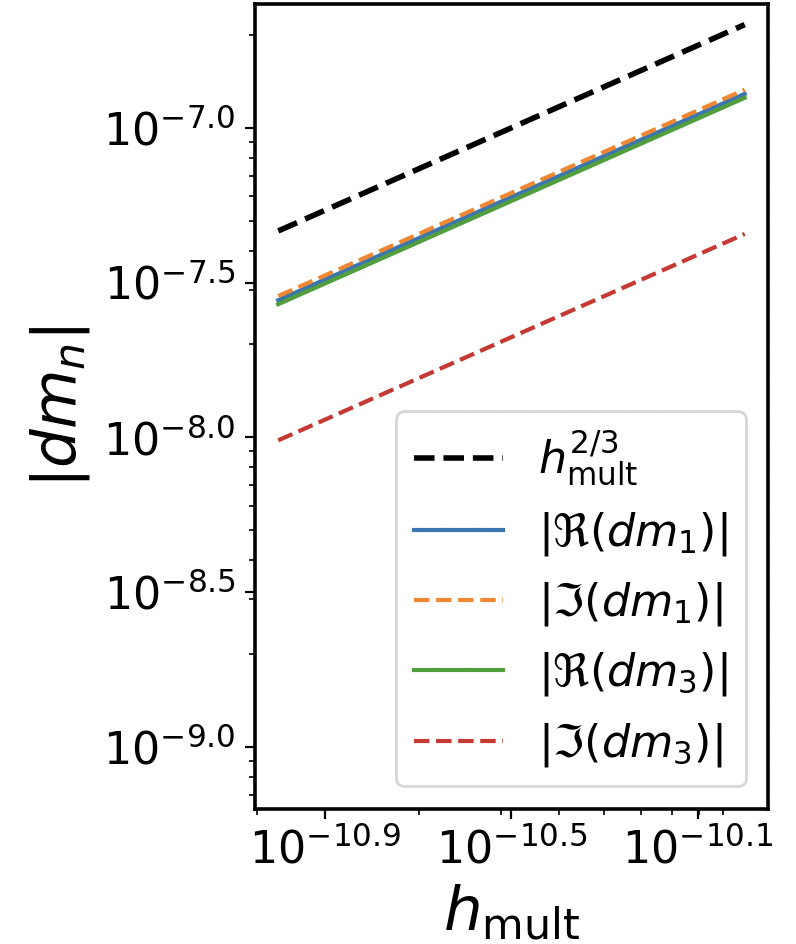}
    \caption{} 
    \label{fig:7b}
  \end{subfigure}
  \caption{Scaling of $\delta m_n$ with $h_{\mathrm{mult}}$ at $P=P_c$:
  (a) even modes with an $h_{\mathrm{mult}}^{1/3}$ reference (dashed black);
  (b) odd modes with an $h_{\mathrm{mult}}^{2/3}$ reference (dashed black). $|\Re\,\delta m_n|$ and $|\Im\,\delta m_n|$ denote real and imaginary parts ($\delta m_n$ is generally complex).}
  \label{fig:dmn_scaling}
\end{figure}

\vspace{-6px}
For each $h_{\mathrm{mult}}$ tested in a narrow logarithmic range, we computed the $P=P_c$ limit cycle,
extracted $m_n$, and formed $\delta m_n=m_n-m_{n,c}$. As shown in Figure~\ref{fig:dmn_scaling}, the
log--log trends in $|\Re\,\delta m_n|$ and $|\Im\,\delta m_n|$ exhibit a parity-dependent
power law:
\[
  |\delta m_{2n}| \sim h_{\mathrm{mult}}^{1/3},\qquad
  |\delta m_{2n+1}| \sim h_{\mathrm{mult}}^{2/3}.
\]\vspace{-15px}

\vspace{-5px}
These exponents are compatible with \(z_k\propto h_{\mathrm{mult}}^{1/3}\):
for even \(k\), \(m_{k,c}=0\) so \(z_k=\sqrt{|m_k|^2}=|\delta m_k|\propto h_{\mathrm{mult}}^{1/3}\), while for odd \(k\),
\[
  z_k^2=\bigl||m_k|^2-|m_{k,c}|^2\bigr|
  \approx 2\,|\Re\!\bigl(m_{k,c}^*\,\delta m_k\bigr)|
  \propto h_{\mathrm{mult}}^{2/3}
  \;\Rightarrow\;
  z_k\propto h_{\mathrm{mult}}^{1/3}.
\]

\vspace{-13px}
\subsection{Odd/even-harmonics and the origin of the parity rule}\vspace{-5px}
To interpret the parity-separated scaling in Figure~\ref{fig:dmn_scaling}, we decomposed the
periodic field and response into odd- and even-harmonic parts,
\(m(t)=m_o(t)+m_e(t)\) and \(h(t)=h_o(t)+h_e(t)\), where \(m_o\) contains only odd Fourier
modes \(m_{2n+1}\) and \(m_e\) contains only even modes \(m_{2n}\) (and similarly for \(h\)).
Separating Eq.~\eqref{eq:tdgl}
into odd and even harmonics gives the coupled system
\begin{align}
  \frac{dm_o}{dt} &= -2a\,m_o - 4b\,m_o^3 - 12b\,m_o m_e^2 + h_o,
  \label{eq:mo_eom}\\
  \frac{dm_e}{dt} &= -2a\,m_e - 4b\,m_e^3 - 12b\,m_o^2 m_e + h_e.
  \label{eq:me_eom}
\end{align}

For a purely odd applied field (\(h_e=0\)), setting \(m_e(t)\equiv 0\) satisfies Eq.~\eqref{eq:me_eom}, leaving a reduced equation for \(m_o(t)\)
\begin{equation}
  \frac{dm_o}{dt} = -2a\,m_o - 4b\,m_o^3 + h_o(t),
  \label{eq:mo_reduced}
\end{equation}
which, when plotted as \(m_o(t)\) versus \(h_o(t)\), yields a symmetric hysteresis loop that loses stability and bifurcates at \(P=P_c\).

At \(P=P_c\), an even drive \(h_e\propto h_{\mathrm{mult}}\) affects the even response via
Eq.~\eqref{eq:me_eom}. Numerically we found \(|\delta m_{2n}|\propto h_{\mathrm{mult}}^{1/3}\) for even modes
(Figure~\ref{fig:dmn_scaling}a). The odd deviations are generated through the coupling term
\(-12b\,m_o m_e^2\) in Eq.~\eqref{eq:mo_eom}, which implies
\(|\delta m_{2n+1}|\propto |m_e|^2 \propto h_{\mathrm{mult}}^{2/3}\), consistent with
Figure~\ref{fig:dmn_scaling}b.
\vspace{-8px}
\subsection{Generality across MFGL free energies}\vspace{-5px}
The same scaling procedures used above, (i) the period scaling
$z_k\sim \varepsilon^{1/2}$ for $P<P_c$, (ii) the conjugate-field scaling
$z_k\sim h_{\mathrm{mult}}^{1/3}$ at $P=P_c$, and (iii) the mode-deviation parity rule
$|\delta m_{2n}|\propto h_{\mathrm{mult}}^{1/3}$ and $|\delta m_{2n+1}|\propto
h_{\mathrm{mult}}^{2/3}$ at $P=P_c$, can be applied to other MFGL models.
For example, a model with $F(m)=a m^2 + b m^6 - h m
  \Rightarrow \frac{dm}{dt} = -2a\,m - 6b\,m^5 + h $
and a mixed form $F(m)=a m^4 + b m^6 - h m
  \Rightarrow \frac{dm}{dt} = -4a\,m^3 - 6b\,m^5 + h $
both exhibited the same scalings $z_k\sim h_{\mathrm{mult}}^{1/3}$ at $P_c$ and 
$z_k\sim \varepsilon^{1/2}$ for $P<P_c$, together with the same even/odd parity 
behavior.
\vspace{-8px}
\section{Numerical methods}
\label{subsec:numerics}\vspace{-3px}
All computations were performed in Python using \texttt{numpy}, \texttt{scipy}, \texttt{matplotlib}, and
\texttt{mpmath}. Following common practice in computational Ginzburg--Landau modeling, we treated all
quantities in Eq.~\eqref{eq:tdgl} as dimensionless ($t$ and $P$ are measured in units of the model’s relaxation timescale);
accordingly, axes in Figs.~1--7 are unitless. To obtain this form, we started from a dimensional free energy
\(F'(m',h')=a'(m')^{2}+b'(m')^{4}-m'h'\) and defined
\(m'=m_{0}m\), \(h'=h_{0}h\), \(F'=F_{0}F\), \(a'=a_{0}a\), \(b'=b_{0}b\).
Choosing scales so that \(a_{0}m_{0}^{2}/F_{0}=b_{0}m_{0}^{4}/F_{0}=m_{0}h_{0}/F_{0}=1\),
we obtained the dimensionless model used in this work.

Hysteresis loops (Fig.~\ref{fig:side-by-side}) were computed with \texttt{solve\_ivp} (RK45) and sampled on uniform grids using
\(t_{\max}=200\), \(\Delta t=1\) for \(P=100\) and \(t_{\max}=20\), \(\Delta t=10^{-3}\) for \(P=1\).
Periodic (limit-cycle) solutions were obtained by a shooting method: we integrated Eq.~\eqref{eq:tdgl} over one period
with \texttt{odeint} (LSODA) on a uniform grid of \(N_{t}=10^{4}\) points (and \(N_{t}=10^{5}\) for Fig.~\ref{fig:dmn_scaling}),
solved \(m(P)-m(0)=0\) with \texttt{fsolve} (\(\texttt{xtol}=10^{-6}\)) to obtain a seed,
and refined using \texttt{mpmath} (\(\texttt{mp.dps}\approx 20\)–25) until \(|m(P)-m(0)|<10^{-10}\)
(using bracketing with \(\texttt{xtol}=10^{-10}\)).

For the even--harmonic perturbation in Sec.~\ref{h_mult-scaling} we used \((h_{0},h_{2},h_{4})=(0.3,0.4,0.2)\).
For the subcritical scaling in Fig.~\ref{fig5} we used five logarithmically spaced values \(\varepsilon\in[10^{-9},10^{-7}]\).
For the mode-coupling reproduction (Fig.~\ref{fig4}) we fixed \(P=5.3193577\), evaluated the \(k=0\) balance in Eq.~\eqref{eq:T123}
using modes \(|k|\le 32\) (with \(n_1,n_2=-16,\dots,16\)), and sampled 50 log-spaced values \(h_0=15\times10^{x}\), \(x\in[-14.2,-11.2]\).

All scripts used for Figs.~1--7 are available at \url{https://github.com/ysatynska/DynamicPhaseTransitions}.
\vspace{-10px}
\section*{Conclusions}
\vspace{-2px}
Using the Mean-Field Ginzburg--Landau model, we clarified conjugate-field and order-parameter definitions for magnetic dynamic phase transitions. For purely odd applied field, we confirmed that the symmetry-broken branch below $P_c$ exhibits $z_k\sim \varepsilon^{1/2}$ across many Fourier modes. At $P=P_c$, the appropriate conjugate field is the even-harmonic part of the applied field: scaling the full even sector by a single multiplier $h_{\mathrm{mult}}$ yields $z_k\sim h_{\mathrm{mult}}^{1/3}$ for $k\le 7$. Mode-resolved deviations obey a robust parity rule, $|\delta m_{2n}|\propto h_{\mathrm{mult}}^{1/3}$ and $|\delta m_{2n+1}|\propto h_{\mathrm{mult}}^{2/3}$, consistent with odd/even-harmonic coupling in the TDGL dynamics. These scalings persist in higher-order MFGL models, suggesting they are generic within this mean-field class.

\bibliography{bibliography}

\end{document}